\documentclass[11pt,a4paper]{article}
\usepackage{amsfonts,amssymb}
\usepackage{cite}
\usepackage[T2A]{fontenc}
\usepackage[cp1251]{inputenc}
\usepackage[english]{babel}
\usepackage{amsmath}%
\usepackage[unicode]{hyperref}

\newtheorem{Theorem}{Theorem}

\def\eq#1{\begin{equation}#1\end{equation}}
\def\matrixx#1{\left(\begin{array}{cc}#1\end{array}\right)}
\def\matrixc#1{\left(\begin{array}{c}#1\end{array}\right)}
\def\diag{\hbox{diag}}
\def\eqs#1{\begin{equation}\begin{split}#1\end{split}\end{equation}}
\def\seqs#1{\begin{equation*}\begin{split}#1\end{split}\end{equation*}}
\def\seq#1{\begin{equation*}#1\end{equation*}}
\def\qed{\vrule height0.6em width0.3em depth0pt}
\font\Sets=msbm10

\def\Z {\hbox{\Sets Z}}
\def\C {\hbox{\Sets C}}

\def\qed{\vrule height0.6em width0.3em depth0pt\medskip}

\title{\bf Peculiar symmetry structure of some known discrete nonautonomous equations}
\author{{\bf R.N. Garifullin$^{1,2}$, I.T. Habibullin$^{1,2}$ and R.I. Yamilov$^1$}
\\$^1$ Ufa Institute of Mathematics, Russian Academy 
of Sciences,\\ 112 Chernyshevsky Street, Ufa 450008, Russian Federation
\\$^2$ Bashkir State University,\\ 32 Zaki Validi Street, Ufa 450074, Russian Federation\\
{\sl E-mails: \url{mailto:rustem@matem.anrb.ru},}\\ {\sl \url{mailto:ismagilhabibullin@gmail.com}, \url{mailto:RvlYamilov@matem.anrb.ru}}
\\{\sl URL: \url{http://matem.anrb.ru/garifullinrn},} \\{\sl \url{http://matem.anrb.ru/habibullinit},  \url{http://matem.anrb.ru/en/yamilovri}}}

\begin{document}
\maketitle

\abstract{We study the generalized symmetry structure of three known discrete nonautonomous equations. One of them is the semidiscrete dressing chain of Shabat. Two others are completely discrete equations defined on the square lattice. The first one is a discrete analogue of the dressing chain introduced by Levi and Yamilov. The second one is a nonautonomous generalization of the potential discrete KdV equation or, in other words, the H1 equation of the well-known Adler-Bobenko-Suris list. We demonstrate that these equations have generalized symmetries in both directions if and only if their coefficients, depending on the discrete variables, are periodic. The order of the simplest generalized symmetry in at least one direction depends on the period and may be arbitrarily high. We substantiate this picture by some theorems in the case of small periods. In case of an arbitrarily large period, we show that it is possible to construct two hierarchies of generalized symmetries and conservation laws. The same picture should take place in case of any nonautonomous equation of the Adler-Bobenko-Suris list.}

\section{Introduction}

We demonstrate that the generalized symmetry structure of some nonautonomous equations may be quite unusual by example of three known equations.  

The first equation reads:
\eq{(u_{n+1,m+1}-u_{n,m})(u_{n+1,m}-u_{n,m+1})=\alpha_n-\beta_m. \label{abs_h1}}  Here $u_{n,m}$ is an unknown function depending on two discrete variables $\ n,m\in \Z$, while $\alpha_n,\beta_m$ are the given functions depending on one discrete variable. Eq. (\ref{abs_h1}) is  the H1 equation of the Adler-Bobenko-Suris list \cite{abs03}. In the autonomous case, it is nothing but the discrete potential KdV equation, which has been known much earlier together with its $L-A$ pair, see e.g. \cite{nc95}.

The second equation is the well-known dressing chain studied, e.g., in \cite{sy90,s92,vs93}:
\eq{\frac{d }{dx}(u_{n}+u_{n+1})=u_{n}^2-u_{n+1}^2+\alpha_{n}-\alpha_{n+1}.\label{dress}} Here the unknown function $u_n=u_n(x)$ depends on one continuous $x$ and one discrete $n$ variables. 
The third equation is a completely discrete analogue of the dressing chain:
\eq{\alpha_n(u_{n,m+1}+1)(u_{n,m}-1)=\alpha_{n+1}(u_{n+1,m+1}-1)(u_{n+1,m}+1),\label{d_dress}} which has been introduced in \cite{ly09}. In the autonomous case see, e.g., \cite{nc95,ht95}.

The equations \eqref{abs_h1} and \eqref{d_dress} belong to the following class of discrete equations on the square lattice:
\begin{equation}F_{n,m}(u_{n+1,m},u_{n,m},u_{n,m+1},u_{n+1,m+1})=0. \label{gF}\end{equation}
In autonomous case, when the function $F_{n,m}$ does not depend on $n,m$ explicitly, all known integrable equations of this form have two hierarchies of generalized symmetries, and this property can be used as a criterion of integrability. 
Generalized symmetries in the $n$-direction have the form
\eq{\frac{d u_{n,m}}{dt_k}=\Phi_{n,m}(u_{n+k,m},u_{n+k-1,m},\ldots,u_{n-k,m}),\label{symk}}
where $k\geq1$, and the number $k$ can be called the order of such symmetry.
Generalized symmetries of an order $l\geq1$ in the $m$-direction have the form
\eq{\frac{d u_{n,m}}{d\tau_l}=\Psi_{n,m}(u_{n,m+l},u_{n,m+l-1},\ldots,u_{n,m-l}).\label{syml}}

In most of autonomous integrable cases, the simplest generalized symmetries in both directions have the orders $k=l=1$, see e.g. \cite{lpsy08,ly11,x09}. These symmetries correspond to integrable Volterra type equations of a complete list obtained in \cite{y83}, see also the review article \cite{y06}. There are a few examples with the simplest symmetries of orders $k=l=2$, see \cite{a11,mx13,shl14}. Up to now there has been known the only example with an essentially asymmetric structure of generalized symmetries. It has been found in \cite{gy12}, see also \cite{gmy14}. In that example, the orders of simplest symmetries are different ($k=2$ and $l=1$), and examples we discuss in this article will be asymmetric in the same sense.

As for the nonautonomous case, the situation is different. We know nonautonomous examples of the form \eqref{gF} with two hierarchies of generalized symmetries \cite{xp09,gy14}. However, there are some known integrable nonautonomous equations which have only one hierarchy of generalized symmetries or have no hierarchy at all. This is the case of nonautonomous equations of the Adler-Bobenko-Suris list. It has been hypothesized in \cite{rh07}, and this will be confirmed in the present paper, that  there is no generalized symmetry in the $n$-direction when $\alpha_n$ is an arbitrary $n$-dependent function and no generalized symmetry in the $m$-direction when $\beta_m$ is an arbitrary $m$-dependent function.

In this paper, instead of arbitrary functions $\alpha_n, \beta_m$ in eq. \eqref{abs_h1}, we consider the concrete ones. We look for functions $\alpha_n, \beta_m$, such that the corresponding equation \eqref{abs_h1} has two hierarchies of generalized symmetries. We prove that symmetries of the form \eqref{symk} and \eqref{syml} exist if and only if $\alpha_n$ and $\beta_m$ are the periodic functions. We do this for some low orders $k$ and $l$ only. The orders $k$ and $l$ of the simplest generalized symmetries \eqref{symk} and \eqref{syml} depend on the periods of $\alpha_n$ and $\beta_m$ and may be arbitrarily high as well as different.

For eq.  \eqref{d_dress} the picture is similar. In case of eq. \eqref{dress}, the form of symmetries is different, but the results are quite analogous too. 

In case of the periodic coefficients $\alpha_n$ and $\beta_m$ in eqs. (\ref{abs_h1},\ref{dress},\ref{d_dress}), whose periods may be arbitrarily large, we demonstrate that two hierarchies of generalized symmetries and conservation laws can be constructed by using known nonautonomous $L-A$ pairs of these equations. We do that, using a method presented in \cite{HY13}.

It seems highly probable that two hierarchies of conservation laws also exist if and only if the coefficients of eqs. (\ref{abs_h1},\ref{dress},\ref{d_dress}) are periodic. This property has been confirmed in a sense in \cite{rh07_1}, where it has been shown for eq. \eqref{abs_h1} that so-called five-point conservation laws disappear when the coefficients $\alpha_n$ and $\beta_m$ become nonconstant.

As a result of our investigation we come to an opinion that eqs. (\ref{abs_h1},\ref{dress},\ref{d_dress}) with the periodic coefficients are ``more integrable''. In this case we can derive, in both directions, generalized symmetries and conservation laws from their $L-A$ pairs, while in general case those $L-A$ pairs seem to be much more inconvenient for use.

In Section \ref{sec_h1} we prove a few theorems showing that two hierarchies of generalized symmetries of eq. \eqref{abs_h1} exist only in the case of periodic coefficients. In particular, we construct an interesting example with a simplest generalized symmetry of the second order in one direction and of the third order in the second one. In Sections \ref{sec_dr},\ref{sec_ddr} we prove analogues theorems for eqs. (\ref{dress},\ref{d_dress}). In Section \ref{sec_theory} we explain how to construct two hierarchies of generalized symmetries and conservation laws for eqs. (\ref{abs_h1},\ref{dress},\ref{d_dress}) with the periodic coefficients. Examples of generalized symmetries of low orders together with their $L-A$ pairs are constructed for such equations in Section \ref{ex_sym}. Conservation laws of low orders are given in Section \ref{sec_claws}. The nature of some generalized symmetries of eqs. (\ref{abs_h1},\ref{dress},\ref{d_dress}) with the periodic coefficients is discussed in Section \ref{sec_sys}.

\section{H1 equation}\label{sec_h1}

We study in this section the H1 equation \eqref{abs_h1} for which we use a natural assumption:
\eq{\alpha_n\neq\beta_m\  \label{h1_con}} for any $ n,m \in \Z$. In opposite case the equation becomes degenerate in some points, see e.g. \cite{ly11} for the autonomous case and \cite{gy14} for the nonautonomous one.

In the autonomous case, generalized symmetries of the Adler-Bobenko-Suris equations and of the H1 equation, in particular, were constructed in \cite{rh07,ttx07,lp07,lps07}. Here we look for generalized symmetries of the H1 equation in the nonautonomous case and obtain, as a result, some statements on the symmetry structure of this equation. 

Due to the invariance of eq. \eqref{abs_h1} with respect to the involution $n\leftrightarrow m, \ \alpha_n\leftrightarrow\beta_m$, we can restrict ourselves to generalized symmetries of the form \eqref{symk}. 
According to its definition, see e.g. \cite{ly09}, the symmetry \eqref{symk} of eq. \eqref{gF} must satisfy the relation
\eqs{\frac{\partial F_{n,m}}{\partial u_{n+1,m}}\Phi_{n+1,m}+\frac{\partial F_{n,m}}{\partial u_{n,m}}\Phi_{n,m}+\frac{\partial F_{n,m}}{\partial u_{n,m+1}}\Phi_{n,m+1}+\frac{\partial F_{n,m}}{\partial u_{n+1,m+1}}\Phi_{n+1,m+1}=0\label{com_con}} on the solutions of eq. \eqref{gF} and for all $n,m\in\Z.$ 

It is natural to assume for the symmetries \eqref{symk} of order $k$ that $\Phi_{n,m}$ depends on both $u_{n+k,m}$ and $u_{n-k,m}$ at at least one point $n,m$. We prove theorems below under a stronger nondegeneracy condition for such generalized symmetries:
\eq{\frac{\partial \Phi_{n,m}}{\partial u_{n+k,m}}\neq0\  \hbox{  and  }\  \frac{\partial \Phi_{n,m}}{\partial u_{n-k,m}}\neq0\ \hbox{ for all}\  n,m\in\Z\label{con_sym}.} 
In all known cases, if an equation of the form \eqref{gF} has a generalized symmetry \eqref{symk} of an order $k\geq1$, then it has the nondegenerate symmetry of the same order. 

We find generalized symmetries by using a scheme developed in \cite{ly11,ggh11}. Some annihilation operators \cite{h05} play an important role in this scheme. 

Eq. (\ref{abs_h1}) has the following point symmetry:
\eqs{\frac{d}{dt_0}u_{n,m}=\nu_1+\nu_2(-1)^{n+m}+\nu_3u_{n,m}(-1)^{n+m},\label{sp}} where $\nu_1,\nu_2,\nu_3$ are arbitrary constants. We write down below generalized symmetries up to this point one.

\subsection{First and second order generalized symmetries}

Here we get some theoretical results in case of the first and second order generalized symmetries. The following result has been obtained in \cite{rh07}, and we present it below for completeness.

\begin{Theorem} Eq. (\ref{abs_h1},\ref{h1_con}) has a first order nondegenerate generalized symmetry in the $n$-direction iff $\alpha_n\equiv\alpha_{n-1}.$\label{th_h1_1}\end{Theorem}

\paragraph{Sketch of proof.} The compatibility condition \eqref{com_con} for eqs. (\ref{abs_h1}) and (\ref{symk}) implies:
\eq{(\alpha_n-\alpha_{n-1})\frac{\partial \Phi_{n,m}}{\partial u_{n+1,m}}\equiv0,\ \  (\alpha_n-\alpha_{n-1})\frac{\partial \Phi_{n,m}}{\partial u_{n-1,m}}\equiv0,\label{con_symp}} and we get the first part of the theorem. On the other hand, eq. (\ref{abs_h1}) with $\alpha_n\equiv\alpha_{n-1}$ has, for any $\beta_m$, the generalized symmetry
\eq{\frac{d}{dt_1}u_{n,m}=\frac1{u_{n+1,m}-u_{n-1,m}}\label{sym_nond}}  which is nondegenerate. \qed

{\bf Remark.} The same result can be obtained under a weaker assumption instead of the nondegeneracy condition \eqref{con_sym} with $k=1$. For example, we can assume that there exists $m$, such that 
\eq{ \frac{\partial \Phi_{n,m}}{\partial u_{n+1,m}}\neq0 \ \hbox{ or }  \ \frac{\partial \Phi_{n,m}}{\partial u_{n-1,m}}\neq0\label{con_syml}} for any $n$. In this case we also derive $\alpha_n\equiv\alpha_{n-1}$ from eq. (\ref{con_symp}).


In the case $\alpha_n\equiv\alpha_{n-1}\equiv\alpha$ and $\beta_{m}\equiv\beta_{m-1}\equiv\beta$, there is one more symmetry:
\eq{\frac{d}{dt'_1}u_{n,m}=\frac{2n(\beta-\alpha)}{u_{n+1,m}-u_{n-1,m}}+u_{n,m}\label{mas_h1_1}.}
Any symmetry of the order $k\leq 1$ of eq. (\ref{abs_h1},\ref{h1_con}) with $\alpha_n\equiv\alpha_{n-1}$ is, up to a point symmetry \eqref{sp}, the following linear combination with constant coefficients $\mu_1,\mu_2$:
\eqs{\frac{du_{n,m}}{dt''_1}=\mu_1\frac{du_{n,m}}{dt_1}+\mu_2\frac{du_{n,m}}{dt'_1}.}

Eq. \eqref{sym_nond} is a known integrable equation of the Volterra type \cite{y83,y06}. Eq. \eqref{mas_h1_1} is its known  master symmetry \cite{asy00}. It generates generalized symmetries not only for eq. \eqref{sym_nond} but also for the discrete equation (\ref{abs_h1},\ref{h1_con}) with $\alpha_n\equiv \alpha$ and an arbitrary $\beta_m$. For example, it can be checked by direct calculation that the following equation, constructed in the standard way, 
\eq{\frac{d}{dt_2}u_{n,m}=\frac{d}{dt_1}\frac{d}{dt'_1}u_{n,m}-\frac{d}{dt'_1}\frac{d}{dt_1}u_{n,m}}
is the second order generalized symmetry for both of these equations.

Below we consider the two-periodic case: $\alpha_{n+1}\equiv\alpha_{n-1}$. There we have two possibilities. First of them is $\alpha_0=\alpha_1$, hence $\alpha_n\equiv\alpha$ is a constant, and we are led to the previous one-periodic case. In the second case $\alpha_0\not=\alpha_{1}$, then $\alpha_{n+1}\not =\alpha_{n}$ for any $n$. 

\begin{Theorem}The following two statements take place:
\begin{enumerate}
\item If eq. (\ref{abs_h1},\ref{h1_con}) has a nondegenerate generalized symmetry of order $2$ in the $n$-direction, then $\alpha_{n+1}\equiv \alpha_{n-1}$. 
\item If $\alpha_n $ in eq. (\ref{abs_h1},\ref{h1_con}) satisfies the conditions \eq{\alpha_{n+1}\equiv \alpha_{n-1} \hbox{ and } \alpha_{0}\neq \alpha_{1}\label{p2n1},} then in the $n$-direction there exists the nondegenerate second order symmetry and there is no symmetry of the first order.
\end{enumerate}
\end{Theorem}

\paragraph {Proof.}
We can derive from the compatibility condition (\ref{com_con}) the following relations:
\seqs{(\alpha_{n+1}-\alpha_{n-1})\frac{\partial \Phi_{n,m}}{\partial u_{n+2,m}}\equiv0,\ \ (\alpha_{n}-\alpha_{n-2})\frac{\partial \Phi_{n,m}}{\partial u_{n-2,m}}\equiv0,\label{con_sym2}} which provide the first part of the theorem. 
In case  \eqref{p2n1} we have  the second order symmetry \eqs{\frac{d}{dt_2}u_{n,m}=\frac{c_n(u_{n,m}-u_{n+2,m})}{\gamma_{n}+(u_{n-1,m}-u_{n+1,m})(u_{n,m}-u_{n+2,m})}\\+\frac{c_{n-1}(u_{n,m}-u_{n-2,m})}{\gamma_{n}+(u_{n-1,m}-u_{n+1,m})(u_{n,m}-u_{n-2,m})},\label{sym_h1_2}} where $c_{n+2}\equiv c_n$ is an arbitrary two-periodic function, and $\gamma_n=\alpha_{n+1}-\alpha_{n}\neq 0$ for any $n$.   This formula yields the nondegenerate symmetries of order 2 (e.g. if $c_n\equiv 1$ or $c_n=2+(-1)^n$).

In the case $\beta_{m}\equiv\beta$ we have an additional symmetry:
\eqs{\frac{d}{dt'_2}u_{n,m}=\frac{n(\beta-\alpha_{n+1})(u_{n,m}-u_{n+2,m})}{\gamma_{n}+(u_{n-1,m}-u_{n+1,m})(u_{n,m}-u_{n+2,m})}\\+\frac{(n-1)(\beta-\alpha_n)(u_{n,m}-u_{n-2,m})}{\gamma_{n}+(u_{n-1,m}-u_{n+1,m})(u_{n,m}-u_{n-2,m})}-u_{n,m}.\label{ms_h1_2}}
Any symmetry of the order $k\leq2$ of eq.(\ref{abs_h1},\ref{h1_con},\ref{p2n1}) is a linear combination of (\ref{sp},\ref{sym_h1_2},\ref{ms_h1_2}). For this reason, there is no first order symmetry. \qed

Let us note that if $\alpha_{n+1}\equiv\alpha_{n}$, then eqs. (\ref{sym_h1_2},\ref{ms_h1_2}) turn into (\ref{sym_nond},\ref{mas_h1_1}).
The formula \eqref{sym_h1_2} provides two linear independent and commuting symmetries of the discrete equation. 

Eq. \eqref{ms_h1_2} should be the master symmetry for \eqref{sym_h1_2}, providing generalized symmetries of even orders not only for \eqref{sym_h1_2} but also for eq. (\ref{abs_h1},\ref{h1_con},\ref{p2n1}) with an arbitrary $\beta_m$. We have checked that by direct calculation in the first step, constructing a fourth order generalized symmetry.

\subsection{An example with asymmetric symmetry structure} \label{asH1}

We consider here eq. (\ref{abs_h1},\ref{h1_con}) satisfying the conditions \eq{\alpha_{n+3}\equiv\alpha_n,\quad \beta_{m+2}\equiv\beta_m\label{p3p2}.} 
We also require: 
\eq{\alpha_0\neq\alpha_1,\quad \alpha_0\neq\alpha_2,\quad \alpha_1\neq\alpha_2,\quad \beta_0\neq\beta_1 \label{alphaa}.}
This provides that
$$\alpha_{n+2}-\alpha_n \equiv \alpha_{n-1}-\alpha_{n}\neq0,\quad \beta_m\neq\beta_{m-1}$$ for all $n,m$.
Taking into account the condition \eqref{h1_con}, we see that all the five numbers $\alpha_0, \alpha_1, \alpha_2, \beta_0, \beta_1$ must be different.

There is in the $n$-direction the following generalized symmetry:
\eqs{\frac{du_{n,m}}{dt_3}=\frac{(v_{n+2,m}v_{n+1,m}+\gamma_{n+1})a_{n+1}}{v_{n+2,m}v_{n+1,m}v_{n,m}-v_{n+2,m}\gamma_{n+2}+v_{n,m}\gamma_{n+1}}\\ +\frac{v_{n+1,m}v_{n-1,m}a_{n}}{v_{n+1,m}v_{n,m}v_{n-1,m}+v_{n+1,m}\gamma_{n+1}+v_{n-1,m}\gamma_n}\\
+\frac{(v_{n-1,m}v_{n-2,m}-\gamma_{n})a_{n-1}}{v_{n,m}v_{n-1,m}v_{n-2,m}-v_{n,m}\gamma_{n}-v_{n-2,m}\gamma_{n-1}},\\ v_{n,m}=u_{n+1,m}-u_{n-1,m},\quad \gamma_n=\alpha_{n+1}-\alpha_n,\quad a_{n+3}\equiv a_{n}.\label{sn32}}
It is of the order $k=3$ and is nondegenerate in particular cases, e.g. $a_n\equiv 1$. There are here three linear independent and commuting generalized symmetries. Any symmetry in the $n$-direction of an order $k\leq 3$ is a linear combination of eqs. \eqref{sn32} and \eqref{sp}. That is why there is no symmetry of the orders $k=1$ and $k=2$.

There is in the $m$-direction the following generalized symmetry of the form (\ref{syml}):
\eqs{\frac{du_{n,m}}{d\tau_2}=\frac{w_{n,m+1}b_{m+1}}{w_{n,m+1}w_{n,m}+\delta_m}
+\frac{w_{n,m-1}b_{m}}{w_{n,m}w_{n,m-1}-\delta_m} ,\\
w_{n,m}=u_{n,m+1}-u_{n,m-1},\quad \delta_m=\beta_{m+1}-\beta_m,\quad b_{m+2}\equiv b_{m}.\label{sm32}}
It is of the order $l=2$ and is nondegenerate in particular cases, e.g. $b_m\equiv 1$. We have here two linear independent and commuting symmetries. Any symmetry in the $m$-direction of an order $l\leq 2$ is a linear combination of eqs. \eqref{sm32} and \eqref{sp}. For this reason there is no symmetry of the order $l=1$.

The results can be formulated as follows:

\begin{Theorem}
Eq. (\ref{abs_h1},\ref{h1_con},\ref{p3p2},\ref{alphaa}) has in the $n$-direction a nondegenerate generalized symmetry of the order $k=3$ and has no symmetry of the orders $k=1,2$. This equation possesses in the $m$-direction a nondegenerate symmetry of the order $l=2$ and has no symmetry of the order $l=1$.
\end{Theorem}

\section{Dressing chain}\label{sec_dr}

In this section we discuss the dressing chain \eqref{dress}.  From the viewpoint of the generalized symmetry properties, equations of the form \eqref{gF} are the discrete analogues of hyperbolic type equations: $$ u_{xy}=f(x,y,u,u_x,u_y).$$ Eq. \eqref{dress} belongs to the class of equations \eq{u_{n+1,x}=f_n(x,u_n,u_{n+1},u_{n,x})\label{sdis}} which are, in the same sense, the semidiscrete analogues of hyperbolic type equations. In the autonomous case, all integrable equations of these three classes should have two hierarchies of generalized symmetries in two different directions. In the paper \cite{y90} a number of autonomous examples of the form \eqref{sdis}, including eq. \eqref{dress} with the constant $\alpha_n$, have been presented together with two generalized symmetries in two different directions.

Eq. \eqref{dress} is a nonautonomous representative of the class \eqref{sdis}. A hierarchy of generalized symmetries in the $x$-direction exists for any $\alpha_n$, and the simplest equation has the form
\eq{\frac{d u_n}{d \theta}=\frac{d^3u_n}{dx^3}-6(u_n^2+\alpha_n)\frac{du_n}{dx}\label{mkdv}.} 
For any fixed $n$ this equation is nothing but the well-known modified Korteweg–de Vries equation.
As it will be shown below, symmetries of eq. \eqref{dress} in the $n$-direction disappear in the generic case, i.e. when $\alpha_n$ is an arbitrary function. We will search the functions $\alpha_n$, such that generalized symmetries in the $n$-direction exist.

Generalized symmetries of eqs. \eqref{sdis} in the  $n$-direction of an order $k\ge 1$ have the form
\eq{\frac{d u_n}{d t_k}=\Phi_n(x,u_{n+k},u_{n+k-1},\dots,u_{n-k}) \label{k-sym}.}
It is natural to assume for such symmetries that $\Phi_{n}$ depends on both $u_{n+k}$ and $u_{n-k}$ in at least one point $n$. We prove theorems below under the following nondegeneracy condition for the generalized symmetries:
\eq{\frac{\partial \Phi_{n}}{\partial u_{n+k}}\neq0\  \hbox{  and  }\  \frac{\partial \Phi_{n}}{\partial u_{n-k}}\neq0\ \hbox{ for all}\  n\in\Z\label{con_sym1}.} 
In all integrable cases we know, if an equation \eqref{sdis} has a generalized symmetry \eqref{k-sym} of an order $k\geq1$, then it  has the nondegenerate symmetry of the same order. 

The generalized symmetry \eqref{k-sym} of eq. \eqref{sdis} must satisfy to the compatibility condition 
\eq{\frac{d}{dx}\Phi_{n+1}=\frac{\partial f_n}{\partial u_{n}}\Phi_n+\frac{\partial f_n}{\partial u_{n+1}}\Phi_{n+1}+\frac{\partial f_n}{\partial u_{n,x}}\frac{d}{dx}\Phi_n\label{con_sdis}}on any solution of eq. \eqref{sdis} and for all $n$.

\subsection{First order generalized symmetries}
\begin{Theorem}Eq. \eqref{dress} has a first order nondegenerate generalized symmetry in the $n$-direction iff $\alpha_n\equiv\alpha_{n+1}$\label{th_dres_1}.\end{Theorem}

\paragraph {Proof.} We can derive from the compatibility condition \eqref{con_sdis} the relations
\seq{(\alpha_{n+1}-\alpha_{n})\frac{\partial \Phi_n}{\partial u_{n+1}}\equiv 0,\quad (\alpha_{n}-\alpha_{n-1})\frac{\partial \Phi_n}{\partial u_{n-1}}\equiv 0} which provide the first part of the theorem. 
On the other hand, eq. \eqref{dress} with $\alpha_n\equiv\alpha_{n+1}$ has the following nondegenerate generalized symmetry:
\eq{\label{sym_dr_1}\frac{du_n}{dt_1}=\frac1{u_{n+1}+u_{n}}-\frac1{u_{n}+u_{n-1}}. \quad\qed }

In the autonomous case, both symmetries (\ref{mkdv}) and \eqref{sym_dr_1} of eq. \eqref{dress} have been found in \cite{y90}.

There is one more generalized symmetry: 
\eq{ \frac{du_n}{dt_1'}=\frac{n}{u_{n+1}+u_{n}}-\frac{n-1}{u_{n}+u_{n-1}}\label{ms_dr_1},} and any symmetry of an order $k\le 1$ of eq. \eqref{dress} with $\alpha_n\equiv\alpha_{n+1}$ is a linear combination of (\ref{sym_dr_1},\ref{ms_dr_1}).
Eq. \eqref{ms_dr_1} provides the master symmetry for eq. \eqref{sym_dr_1}. For example,
\eqs{\frac{du_n}{dt_2}&=\left(\frac{d}{dt_1}\frac{d}{dt'_1}-\frac{d}{dt'_1}\frac{d}{dt_1}\right)u_n=\frac{1}{(u_{n+2}+u_{n+1})(u_{n+1}+u_{n})^2}\\&-\frac{u_{n+1}-u_{n-1}}{(u_{n+1}+u_{n})^2(u_{n}+u_{n-1})^2}-\frac{1}{(u_{n}+u_{n-1})^2(u_{n-1}+u_{n-2})}} is the generalized symmetry not only for eq. \eqref{sym_dr_1} but also for the autonomous semidiscrete equation \eqref{dress}.

\subsection{Second order generalized symmetries}\label{sec_or_dr}

\begin{Theorem}
The following two statements take place:
\begin{enumerate}
\item If eq. \eqref{dress} has a nondegenerate generalized symmetry of the second order in the $n$-direction, then $\alpha_{n+1}\equiv \alpha_{n-1}$. 
\item If $\alpha_n $ of eq. (\ref{dress}) satisfies the conditions  \eq{\alpha_{n+1}\equiv \alpha_{n-1} \hbox{ and } \alpha_{0}\neq \alpha_{1},\label{p2n11}}  then there exists in the $n$-direction  a nondegenerate second order symmetry  and there is no symmetry of the first order.
\end{enumerate}
\end{Theorem}

\paragraph {Proof.}
The compatibility condition \eqref{con_sdis} implies
\seqs{(\alpha_{n+2}-\alpha_{n})\frac{\partial \Phi_{n}}{\partial u_{n+2}}\equiv0,\ \  (\alpha_{n}-\alpha_{n-2})\frac{\partial \Phi_{n}}{\partial u_{n-2}}\equiv0\label{con_sym2p},} and we get the first part of the theorem. 

In the case \eqref{p2n11} all symmetries of orders $k\leq2$ are described as follows: \eqs{\frac{d u_{n}}{dt_2}=&\frac{a_{n+2}(u_{n+2}+u_{n+1})}{\gamma_{n}+(u_{n+2}+u_{n+1})(u_{n+1}+u_{n})}\\+&\frac{a_{n+1}(u_{n+1}-u_{n-1})}{\gamma_{n}-(u_{n+1}+u_{n})(u_{n}+u_{n-1})}
-\frac{a_{n}(u_{n-1}+u_{n-2})}{\gamma_{n}+(u_{n}+u_{n-1})(u_{n-1}+u_{n-2})},\label{sym_dr_2}} where $\gamma_n=\alpha_{n+1}-\alpha_{n}\neq 0$  for all $n$. The function $a_n$ is given by $a_n=b_n+cn,$ where $b_n$ is an arbitrary two-periodic function and $c$ is an arbitrary constant.

There are here nondegenerate examples of the second order, e.g. $a_n\equiv 1$ or $a_n=2+(-1)^n$, but there is no symmetry of the first order. \qed

In the case when $a_n$ is the two-periodic function, i.e. $a_n=b_n$, we have in (\ref{sym_dr_2}) two linear independent and commuting generalized symmetries.
The linear case $a_n=cn$ provides the master symmetry for  eq. \eqref{sym_dr_2} with $a_n=b_n$. This master symmetry  generates symmetries not only for eq. \eqref{sym_dr_2} with $a_n=b_n$ but also for the semidiscrete equation (\ref{dress},\ref{p2n11}). We have checked that on the first step, constructing a fourth order generalized symmetry.

\section{Discrete dressing chain}\label{sec_ddr}
In this section we discuss eq. \eqref{d_dress} with $\alpha_n\neq0$ for any $n$. In \cite{ly09} a complete analogue of the dressing chain \eqref{dress} has been introduced in the following form:
\eq{(v_{n+1,m}+d_m)(v_{n,m}-d_m)=(v_{n+1,m+1}-d_{m+1})(v_{n,m+1}+d_{m+1}).\label{d_dress08}} 
If $d_m\neq0$ for all $m$, then after using the involution $n\leftrightarrow m$ and an obvious rescaling of $v_{n,m}$ we obtain the discrete equation \eqref{d_dress} with $\alpha_n=d_n^2\neq0$ for any $n$. This form is more comfortable for further investigation.

For any $\alpha_n$ there exists a hierarchy of generalized symmetries of eq. \eqref{d_dress} in the $m$-direction, and its simplest representative reads \cite{ly09}:
\eq{\frac{d u_{n,m}}{d \tau_1}=\alpha_n(u_{n,m}^2-1)(u_{n,m+1}-u_{n,m-1})\label{mVol}.} For any fixed $n$ it obviously is the modified Volterra equation. We will look for functions $\alpha_n$, such that generalized symmetries in the $n$-direction exist. 

It should be remarked that an integrable generalization of eq. \eqref{d_dress} has been presented in \cite{gy14} together with one hierarchy of generalized symmetries and an $L-A$ pair.

\subsection{Simplest case}

The following result has been obtained in \cite{gy14}, and we present it here for completeness.

\begin{Theorem}Eq. \eqref{d_dress} with $\alpha_n\neq0$ for any $n$ has a first order nondegenerate generalized symmetry of the form \eqref{symk} iff $\alpha_n\equiv\alpha_{n-1}.$\label{th_dd_1}\end{Theorem}

In the case $\alpha_n\equiv\alpha_{n-1}$, the following nondegenerate symmetry is known \cite{ly09}:
\eq{\frac{d u_{n,m}}{dt_1}=(u_{n,m}^2-1)\left(\frac{1}{u_{n+1,m}+u_{n,m}}-\frac{1}{u_{n,m}+u_{n-1,m}}\right).\label{sym_dd}}
We just can add that there is one more generalized symmetry in the $n$-direction:
\eq{\frac{d u_{n,m}}{dt'_1}=(u_{n,m}^2-1)\left(\frac{n}{u_{n+1,m}+u_{n,m}}-\frac{n-1}{u_{n,m}+u_{n-1,m}}\right),\label{ms_dd}} 
and any symmetry of an order $k\leq1$ of eq. \eqref{d_dress} is a linear combination of \eqref{sym_dd} and \eqref{ms_dd}.

Eq. \eqref{ms_dd} is the known master symmetry of eq. \eqref{sym_dd} \cite{cy95}. It provides generalized symmetries not only for \eqref{sym_dd} but also for eq. \eqref{d_dress}. 

\subsection{Second order generalized symmetries}

\begin{Theorem}
The following two statements take place:
\begin{enumerate}
\item If eq. \eqref{d_dress} with $\alpha_n\neq0$ for all $n$ has a nondegenerate generalized symmetry of the second order in the $n$-direction, then $\alpha_{n+1}\equiv \alpha_{n-1}$. 
\item If $\alpha_n\neq0 $ for any $n$ in eq. (\ref{d_dress}) and it satisfies the conditions  \eq{\alpha_{n+1}\equiv \alpha_{n-1} \hbox{ and } \alpha_{0}\neq \alpha_{1},\label{p2n13}}  then there exists in the $n$-direction a nondegenerate second order symmetry of eq. (\ref{d_dress}) and there is no symmetry of the first order.
\end{enumerate}
\end{Theorem}

\paragraph {Proof.}
The compatibility condition \eqref{com_con} implies:
\seqs{(\alpha_{n+2}-\alpha_{n})\frac{\partial \Phi_{n,m}}{\partial u_{n+2,m}}\equiv0,\ \  (\alpha_{n}-\alpha_{n-2})\frac{\partial \Phi_{n,m}}{\partial u_{n-2,m}}\equiv0,\label{con_sym2l}} and we get the first part of the theorem. 

In the case  \eqref{p2n13}, all symmetries of orders $k\leq2$ in the $n$-direction are described as follows: \eqs{\frac{d u_{n,m}}{dt_2}=&(u_{n,m}^2-1)\left(\frac{a_{n+2}(u_{n+2,m}+u_{n+1,m})}{U_{n+1,m}}\right.\\
-&\left.\frac{a_{n+1}(u_{n+1,m}-u_{n-1,m})}{U_{n,m}}-\frac{a_{n}(u_{n-1,m}+u_{n-2,m})}{U_{n-1,m}}\right),\label{sym_ddr_2}\\ U_{n,m}=&(u_{n+1,m}+u_{n,m})(u_{n,m}+u_{n-1,m})+(u_{n,m}^2-1)(\beta_{n-1}-1),} where $\beta_n=\alpha_{n+1}/\alpha_n\neq1$ for all $n$. The function $a_n$ is given by $a_n=b_n+cn$, where $b_n$ is an arbitrary two-periodic function and $c$ is an arbitrary constant.

There are in (\ref{sym_ddr_2}) nondegenerate examples of the second order, e.g. $a_n\equiv 1$ or $a_n=2+(-1)^n$, but there is no symmetry of the first order. \qed

In the case when $a_n$ is the two-periodic function, i.e. $a_n=b_n$, we have in (\ref{sym_ddr_2}) two linear independent and commuting generalized symmetries.
The linear case $a_n=cn$ provides the master symmetry for  eq. \eqref{sym_ddr_2} with $a_n=b_n$, which generates generalized symmetries for this equation. Those generalized symmetries should be the symmetries of the discrete equation (\ref{d_dress},\ref{p2n13}) too, as the Lie algebra of symmetries should be closed under the operation of commutation, but the verification of this property is difficult even on the first step.

\section{Method of the construction of generalized symmetries and conservation laws }\label{sec_theory}

A method has been developed in \cite{HY13} for the autonomous and weakly nonautonomous discrete and semidiscrete equations, which allows one to construct generalized symmetries and conservation laws by using the $L-A$ pairs.
That method is based on the formal diagonalization of an $L-A$ pair in the neighborhood of a stationary singular point. In this section we generalize that method to the case of the nonautonomous equations with periodic coefficients.

\subsection{Formal diagonalization}
Let us first discuss the formal diagonalization in the case of systems of the linear discrete equations.

We consider a discrete linear vector equation of the form
\begin{equation}\label{linearequation}
\Psi_{n+k}=f_n({\bf u}_n,\lambda)\Psi_n, \quad k\geq1,
\end{equation}
where $\Psi_n$ is an unknown vector, the matrix potential $f_n({\bf u}_n,\lambda)\in {\C^{s\times s}}$ is a meromorphic function of $\lambda\in{\C}$, and the vector function ${\bf u}_n$ is a functional parameter. A point $\lambda=\lambda_0$ is called the point of singularity of eq. (\ref{linearequation}) if it is either a pole of $f_n({\bf u}_n,\lambda)$ or a solution of the equation $\det f_n({\bf u}_n,\lambda)=0$. It is assumed that the set of roots of the equation $\det f_n({\bf u}_n,\lambda)=0$, as well as the set of poles of $f_n$, does not depend on $n$.

We suppose here that eq. (\ref{linearequation}) with the singular point $\lambda_0$ is reduced to the following special form:
\begin{equation}\label{lin_equation2}
\Psi_{n+k}=P_n({\bf u}_n,\lambda)Z\Psi_n,
\end{equation}
where $Z$ is a diagonal matrix $Z=\diag((\lambda-\lambda_0)^{\gamma_1}, (\lambda-\lambda_0)^{\gamma_2},\dots , (\lambda-\lambda_0)^{\gamma_s})$ with integer exponents $\gamma_j$, such that
$\gamma_1<\gamma_2<... <\gamma_s.$

It should be remarked that there is no proof that eq. \eqref{linearequation} can be transformed into the form \eqref{lin_equation2}. However, there is a general scheme which provides, as a rule, a transition from eq. \eqref{linearequation} to \eqref{lin_equation2}. That scheme has been presented in \cite{HY13}. Besides, it will be explained in detail in Section \ref{s_3_ex} for three examples under consideration how to get the representation \eqref{lin_equation2}.

Let us rewrite eq. (\ref{lin_equation2}) as $L\Psi_n=\Psi_n$ with 
\begin{equation}\label{dis_oper}
L=D_n^{-k}P_n({\bf u}_n,\lambda)Z,
\end{equation}
where $D_n$ is the shift operator acting by the rule $D_n:\, n\rightarrow\, {n+1}$. The following statement on the formal diagonalization of the operator $L$ takes place, see \cite{HY13,H85}.

\begin{Theorem}\label{diagon}
Assume that, for any integer $n$ and for ${\bf u}_n$ ranging in a domain, the function $P_n({\bf u}_n,\lambda)$ is analytic in a neighborhood of the point $\lambda_0$, and all the leading principal minors ${\det_j}$ of the matrix $P_n({\bf u}_n,\lambda_0)$ do not vanish:
$$\det_jP_n({\bf u}_n,\lambda_0)\neq0 \quad\mbox{for }\quad j=1,2,...,s\quad \mbox{and for all}\quad n.$$
Then there exists a formal series $T_n=\sum_{i\geq0} T_n^{(i)}(\lambda-\lambda_0)^i$, $\det T_n^{(0)}\neq0$, with the matrix coefficients, such that the operator $L_0=T^{-1}_nLT_n$ is of the form $L_0=D_n^{-k}h_nZ$, where $h_n=\sum_{i\geq0} h^{(i)}_n(\lambda-\lambda_0)^i$ is a formal series with the diagonal coefficients $h_n^{(i)}$, $\det h_n^{(0)}\neq0$.

The series $T_n$ is defined up to multiplication by a formal series with the diagonal coefficients. The latter can be chosen so that all the coefficients $T_n^{(i)}$ and $h_n^{(i)}$ depend on some finite sets of dynamical variables in $\left\{{\bf u}_p\right\}^{p=\infty}_{p=-\infty}$, which in turn depend on $i$.
\end{Theorem}

For $k=1$ Theorem \ref{diagon} has been proved in \cite{HY13}. In general case $k>1$, it can be proved by almost verbatim repeating a proof of \cite{HY13}. It should be remarked that there is in that proof an algorithm for recurrent construction of the coefficients $T_n^{(i)}$, $h_n^{(i)}$.

\subsection[{Diagonalization of the L-A pair and construction of conservation laws}]{Diagonalization of the $L-A$ pair and construction of conservation laws}\label{sec_d_law}
In this section we apply Theorem \ref{diagon} to operators defining the $L-A$ pairs of discrete or semidiscrete scalar equations like eqs. (\ref{abs_h1},\ref{dress}). We also explain how to derive a hierarchy of conservation laws from so-diagonalized operators. The same procedure can be used for analogous systems of discrete or semidiscrete equations.

First we consider a discrete equation \eqref{gF} and suppose that it
is represented as the consistency condition of the following system of linear discrete equations:
\begin{equation}\label{lin_eq3}
\Psi_{n+p,m}=P_{n,m}([{ u}_{n,m}],\lambda)Z\Psi_{n,m}, \quad \Psi_{n,m+q}=R_{n,m}([{ u}_{n,m}],\lambda)\Psi_{n,m}.
\end{equation}
Here the symbol $[{ u}_{n,m}]$ indicates that the matrix functions $P_{n,m}$ and $R_{n,m}$ depend on the dynamical variable ${ u}_{n,m}$ and on a finite number of its shifts. Note that the first equation in (\ref{lin_eq3}) is of the form (\ref{lin_equation2}).
Let us suppose that it satisfies all the conditions of Theorem \ref{diagon}. Then, due to the theorem, the operator $L=D_n^{-p}P_{n,m}Z$ is diagonalized by the conjugation by an appropriate formal series $T_{n,m}$. We assume that the potential $R_{n,m}([{ u}_{n,m}],\lambda)$ rationally depends on $\lambda$. Evidently, the consistency condition for the system (\ref{lin_eq3}) is equivalent to the commutativity condition for the operators $L$ and $M=D_m^{-q}R_{n,m}$. 

It has been proved in \cite{HY13} that the operator $M$ commuting with $L$ is diagonalized by the conjugation with the same series $T_{n,m}$. Therefore we have a diagonal operator $M_0$ with the following representation as a formal series:
\seqs{&M_0=T_{n,m}^{-1}MT_{n,m}=D_m^{-q}S_{n,m}, \\ &S_{n,m}=(\lambda-\lambda_0)^{q_0}(S_{n,m}^{(0)}+S_{n,m}^{(1)}(\lambda-\lambda_0)+ S_{n,m}^{(2)}(\lambda-\lambda_0)^2+\dots ).}
The commutativity condition $[L_0,M_0]=0$, where $L_0=T_{n,m}^{-1}LT_{n,m} = D_n^{-p}h_{n,m}Z$, gives rise to the equation 
\begin{equation}\label{cons_laws2}
h_{n,m+q}S_{n,m}=S_{n+p,m}h_{n,m}, 
\end{equation}
as $[Z,S_{n,m}]=[Z,h_{n,m}]=0$. Eq. (\ref{cons_laws2}) implies the relation $(D^q_m-1)\log h_{n,m}=(D^p_n-1)\log S_{n,m}$. Here and below the notations $\log h_{n,m}$, $\log S_{n,m}$ mean that we apply the logarithm to coefficients of the diagonal matrices.  Therefore the matrix function $H=(D^{q-1}_m+D^{q-2}_m+\dots +1)\log h_{n,m}$ is the generating function for conservation laws.
In this way we get an infinite sequence of conservation laws for the discrete equation (\ref{gF}) whenever the system (\ref{lin_eq3}) satisfies the conditions of Theorem \ref{diagon}.

In a similar way one can consider the semidiscrete model \eqref{sdis}.
Assume that the equation admits an $L-A$ pair of the form
\begin{equation}\label{lin_eq4}
\Psi_{n+k}=P_n([{ u_n}],\lambda)Z\Psi_n, \quad \Psi_{n,x}=A_n([{ u_n}],\lambda)\Psi_n.
\end{equation}
Here the symbol $[{ u}_n]$ indicates the dependence on the dynamical variable ${ u}_{n}$ and on a finite number of its shifts and $x$-derivatives. Let us suppose that the first of eqs. (\ref{lin_eq4}) satisfies the conditions of Theorem \ref{diagon}. 
The compatibility condition for eqs. \eqref{lin_eq4} takes the form $[L,D_x-A_n]=0$, where $L$ is given by (\ref{dis_oper}), and it is equivalent to the semidiscrete equation \eqref{sdis}.
  According to Theorem \ref{diagon} the operator $L$ is diagonalized by the conjugation transform $L_0=T_n^{-1}LT_n$. As it has been proved in \cite{HY13}, the second operator $D_x-A_n$ is diagonalized by the same conjugation $D_x-B_n=T_n^{-1}(D_x-A_n)T_n$, where $B_n=-T_n^{-1}T_{n,x}+T_n^{-1}A_nT_n$ is a formal series with the diagonal coefficients. The commutativity condition $[L_0,D_x-B_n]=0$ of the diagonal operators implies an equation of the form
\begin{equation}\label{cons_laws3}
D_x\log h_n=(D_n^k-1)B_n,
\end{equation}
which generates an infinite sequence of conservation laws. The diagonal operator $h_n$ is defined in Theorem \ref{diagon}.

\subsection{Three examples}\label{s_3_ex}
Here we apply the above method to nonautonomous discrete and semidiscrete models with periodic coefficients. 
We show how to derive from the known $L-A$ pairs of eqs. (\ref{abs_h1},\ref{dress},\ref{d_dress}) the representation \eqref{lin_equation2}. We also check that the conditions of Theorem \ref{diagon} are satisfied.

\subsubsection{H1 equation}
We consider eq. (\ref{abs_h1},\ref{h1_con}) satisfying the restrictions
\begin{equation}\label{per_h1}
\alpha_n\equiv\alpha_{n+N},\quad \beta_m\equiv\beta_{m+K} ,
\end{equation}
\eqs{\label{noteq}&\alpha_k\neq\alpha_{l}\ \hbox{ for } \ 0\leq k<l\leq N-1,\\ &\beta_k\neq\beta_{l} \ \hbox{ for } \ 0\leq k<l\leq K-1.}
An $L-A$ pair for this equation is known \cite{abs03}, see also \cite{nc95} for the autonomous case. It can be written in the form
\eq{\Psi_{n+1,m}=L^{(1)}_{n,m}\Psi_{n,m},\quad \Psi_{n,m+1}=L^{(2)}_{n,m}\Psi_{n,m}\label{lax_h1},} where
\eq{L^{(1)}_{n,m}=\matrixx{-u_{n+1,m}&-1\\ u_{n,m}u_{n+1,m}+\alpha_n-\lambda&u_{n,m}},\quad L^{(2)}_{n,m}=\matrixx{-u_{n,m+1}&-1\\ u_{n,m}u_{n,m+1}+\beta_m-\lambda&u_{n,m}}.}

Theorem~\ref{diagon} cannot be applied directly to any of the linear equations \eqref{lax_h1}
because their singularity points $\lambda=\alpha_n$ and $\lambda=\beta_m$ vary with the discrete variables $n,m$. However, instead of the $L-A$ pair (\ref{lax_h1}), one can use a compound one, for instance: 
\eqs{\Psi_{n+N,m}=f_{n,m}\Psi_{n,m},\quad \Psi_{n,m+1}=L^{(2)}_{n,m}\Psi_{n,m}\label{lax_h1_N},}
with the new potential 
$$f_{n,m}=L^{(1)}_{n+N-1,m}L^{(1)}_{n+N-2,m}\dots L^{(1)}_{n,m}.$$
It has the singularity set $\{\infty, \alpha_0, \alpha_{1},\dots,\alpha_{N-1}\}$ which does not depend on the variables $n,m$. 

Let us transform the first of eqs. (\ref{lax_h1_N}) into the special form (\ref{lin_equation2}). To this end we factorize the matrix $L^{(1)}_{n,m}$ as follows:
$$L^{(1)}_{n,m}=\delta_{n,m}Z\rho_{n,m},$$
where $Z=\diag(1,\lambda-\alpha_n)$ is the diagonal matrix and $\delta_{n,m},\,\rho_{n,m}$ have a triangular structure:
$$\delta_{n,m}=\matrixx{1&0\\ -u_{n,m}+\frac{\lambda-\alpha_n}{u_{n+1,m}}&-\frac{1}{u_{n+1,m}}}, \quad \rho_{n,m}=\matrixx{-u_{n+1,m}&-1\\ 0&1}.$$
Then we change the unknown vector function: $\Psi_{n,m}=\rho^{-1}_{n,m}\Phi_{n,m}$. As a result eqs. \eqref{lax_h1_N} take the form
\eqs{\label{lax_h1_NP}\Phi_{n+N,m}=P_{n,m}Z\Phi_{n,m},\quad \Phi_{n,m+1}=R_{n,m}\Phi_{n,m},  }
where $$P_{n,m}=\rho_{n+N,m}  L^{(1)}_{n+N-1,m}L^{(1)}_{n+N-2,m}\dots L^{(1)}_{n+1,m} \delta_{n,m},\quad R_{n,m}=\rho_{n,m+1}L^{(2)}_{n,m}\rho^{-1}_{n,m}.$$

{\bf Important remark.}
In Theorem \ref{diagon} the singular point $\lambda_0$ should be a constant. In eqs. \eqref{lax_h1_NP} the singular point $\alpha_n$ is not constant, but it is invariant under the action of the shift $D_n^{-N}$. This property also enables us to apply the theorem and to diagonalize the $L-A$ pair \eqref{lax_h1_NP}.

Let us explain why the first equation of the $L-A$ pair \eqref{lax_h1_NP} satisfies the conditions of Theorem \ref{diagon}.
Evidently $P_{n,m}(\lambda)$ is a polynomial of $\lambda$ and therefore it is analytical around the point $\lambda=\alpha_n$. Its determinant is explicitly evaluated: 
$$\det P_{n,m}(\lambda)=\frac{u_{n+N+1,m}}{u_{n+1,m}}(\lambda-\alpha_{n+1})(\lambda-\alpha_{n+2})\dots (\lambda-\alpha_{n+N-1}).$$ At the point $\lambda=\alpha_n$ it is correctly defined and different from zero if the restrictions \eqref{noteq} are  valid  and $u_{n,m}\neq0 $ for all $n,m$. 

The leading principal minor $\det_1 P_{n,m}({\bf u},\lambda=\alpha_n)$, which is located at the left upper corner of the matrix, is a rational function of the coordinates of the point ${\bf u}=(u_{n,m},u_{n+1,m},...,u_{n+N+1,m})\in{\bf C}^{N+2}$, i.e. it is a ratio of two polynomials $Q_1({\bf u}) / Q_2({\bf u})$. Thus there are two possibilities:  $Q_1({\bf u})\equiv0$, and then $\det_1 P_{n,m}( {\bf u},\lambda=\alpha_n)\equiv0$, or $Q_1({\bf u})$ is not identically zero. The first case is not realized, as the leading principal minor under consideration does not vanish at ${\bf u}_{0}=(1,0,0,...,0,1)$. This is easily seen from the following  explicit formula:
$$\det_1 P_{n,m}({\bf u}_{0},\lambda=\alpha_n)=(-1)^N\prod^{[N/2]}_{j=1}(\alpha_n-\alpha_{n+2j-1}).$$
So the polynomial $Q_1({\bf u})$ is nontrivial, and therefore a complete set of its zeros constitutes a manifold $M_1$ of a dimension not greater than $N+1$. Let us denote by $M_2$ the set of zeros of the denominator $Q_2({\bf u})$. Then the function $\det_1 P_{n,m}({\bf u},\lambda=\alpha_n)$ is defined and different from zero in the open domain $\C^{N+2}\backslash(M_1\cup M_2).$

Hence the first equation of the $L-A$ pair satisfies the conditions of Theorem \ref{diagon}. We have proved that if $\alpha_n$ satisfies (\ref{per_h1},\ref{noteq}) and $\beta_m$ is an arbitrary function, then the equation  (\ref{abs_h1},\ref{h1_con}) admits an infinite sequence of conservation laws. In a similar way one can prove that eq.  (\ref{abs_h1},\ref{h1_con}) with an arbitrary $\alpha_n$ and $\beta_m$ satisfying (\ref{per_h1},\ref{noteq}) also possesses an infinite sequence of conservation laws. Certainly, if the restrictions (\ref{per_h1},\ref{noteq}) are true for both $\alpha_n$ and $\beta_m$, then the equation admits two different hierarchies of  conservation laws.

\subsubsection{Dressing  chain}
The second example is the dressing chain (\ref{dress}) obeying the constraint
\begin{equation}\label{per_dress}
\alpha_n\equiv\alpha_{n+N},\quad \mbox{such that} \quad \alpha_k\neq\alpha_{l} \quad \mbox{for} \quad 0\leq k<l\leq N-1.
\end{equation}
Recall that the dressing chain is the compatibility condition \cite{sy90} of the following system of equations:
\eq{\Psi_{n+1}=L^{(1)}_{n}\Psi_{n},\quad D_x\Psi_{n}=Y_{n}\Psi_{n},\label{lax_dress}}
with the potentials
\eq{L^{(1)}_{n}=\matrixx{-u_{n}&1\\ u_{n}^2+\alpha_n+\lambda&-u_{n}},\quad Y_{n}=\matrixx{0&1\\u_n^2+u_{n,x}+\alpha_n+\lambda&0}.}
As this equation is of the hyperbolic type, it may have two hierarchies of conservation laws. One of them has been found in \cite{HY13} by diagonalizing the second of eqs. (\ref{lax_dress}) around the singular point $\lambda=\infty$ and without imposing on $\alpha_n$ any restriction. 

In order to find the second hierarchy, we have to use the first equation of (\ref{lax_dress}). This part of the problem is much more complicated because the singular point $\lambda=-\alpha_n$ depends on $n$. In our opinion, the second hierarchy does exist only under an additional constraint, for instance, (\ref{per_dress}). In case of \eqref{per_dress}, one can avoid difficulties by passing from (\ref{lax_dress}) to a combined Lax pair:
\eq{\Psi_{n+N}=f_{n}\Psi_{n},\quad \Psi_{n,x}=Y_{n}\Psi_{n},\label{lax_dress_N}}
with the potential 
$$f_{n}=L^{(1)}_{n+N-1}L^{(1)}_{n+N-2}\dots L^{(1)}_{n}.$$
Here the potential $f_n$ has the set of singularity points $\{\infty, -\alpha_{0},-\alpha_{1},\dots ,-\alpha_{N-1}\},$ which does not depend on $n$ due to the periodicity condition (\ref{per_dress}). 

Let transform  the first of eqs. (\ref{lax_dress_N}) into the special form (\ref{lin_equation2}). We first factorize the matrix $L_n^{(1)}$ as follows:
$$L^{(1)}_{n}=\delta_{n}Z\rho_{n},$$
where $Z=\diag(1,\lambda+\alpha_n)$ and
$$\delta_{n}=\matrixx{1&0\\ -u_{n}-\frac{\lambda+\alpha_n}{u_{n}}&-\frac{1}{u_{n}}}, \quad \rho_{n}=\matrixx{-u_{n}&1\\ 0&-1}.$$
Then, by  changing the unknown vector function: $\Psi_{n}=\rho^{-1}_{n}\Phi_n$, one rewrites the linear system (\ref{lax_dress_N}) in the following special form:
\eq{\Phi_{n+N}=P_{n}Z\Phi_{n},\quad \Phi_{n,x}=R_{n}\Phi_{n},}  \label{lax_dress_NP}
where $P_{n}=\rho_{n+N}  L^{(1)}_{n+N-1}L^{(1)}_{n+N-2}\dots L^{(1)}_{n+1} \delta_{n}$ and $R_{n}=\rho_{n,x}\rho^{-1}_{n}+\rho_{n}Y_{n}\rho^{-1}_{n}$.
 
The matrix valued function $P_n(\lambda)$ is analytic around the point $\lambda=-\alpha_n$, and its determinant evaluated at $\lambda=-\alpha_n$ is easily found:
$$\det P_n(-\alpha_n)=\frac{u_{n}}{u_{n+N}}(\alpha_n-\alpha_{n+N-1})(\alpha_n-\alpha_{n+N-2})\dots (\alpha_n-\alpha_{n+1}).$$
It is correctly defined and different from zero if $u_n\neq0$ for all $n$.
The leading principal minor $\det_1P_n({\bf u},-\alpha_n)$ is a rational function of the $N+1$-dimensional  variable ${\bf u}=(u_n,u_{n+1},...,u_{n+N})$. We evaluate $\det_1P_n({\bf u}_0,-\alpha_n)$ at ${\bf u}_0=(u_n,0,...,0,u_{n+N})$ and get
$$\det_1P_n({\bf u}_0,-\alpha_n)=-u_{n+N}\prod_{j=1}^{k}(\alpha_{n+2j-1}-\alpha_n)-u_{n}\prod_{j=1}^{k}(\alpha_{n+2j}-\alpha_n)\quad \mbox{if}\quad N=2k+1$$ or
$$\det_1P_n({\bf u}_0,-\alpha_n)=\prod_{j=1}^{k}(\alpha_{n+2j-1}-\alpha_n)+u_{n}u_{n+N}\prod_{j=1}^{k-1}(\alpha_{n+2j}-\alpha_n)\quad \mbox{if}\quad N=2k.$$
The last formulas convince us that the rational function $\det_1P_n({\bf u},-\alpha_n)$ does not equal to zero identically. Hence there is an open domain in ${\C}^{N+1}$ in which $\det_1P_n({\bf u},-\alpha_n)$ does not vanish. So, all the conditions of Theorem \ref{diagon} are satisfied, and the dressing chain (\ref{dress}) with the periodic coefficients (\ref{per_dress}) admits the second hierarchy of conservation laws.

\subsubsection{Discrete dressing  chain}
The third example is the discrete dressing chain (\ref{d_dress}) admitting the following Lax pair \cite{ly09}:
\eq{\Psi_{n+1,m}=L^{(1)}_{n,m}\Psi_{n,m},\quad \Psi_{n,m+1}=L^{(2)}_{n,m}\Psi_{n,m},\label{lax_d_dress}}
\eqs{&L^{(1)}_{n,m}=\matrixx{1&2\lambda\alpha_n(u_{n,m}+1)\\ -\frac2{u_{n,m}-1}&\frac{u_{n,m}+1}{u_{n,m}-1}},\\ &L^{(2)}_{n,m}=\matrixx{1&-\lambda\alpha_n(u_{n,m}+1)(u_{n,m+1}-1)\\ 1&0}.}
Let us assume that the same periodicity constraint (\ref{per_dress}) on $\alpha_n$ is imposed. We transform the $L-A$ pair as:
\eq{\Psi_{n+N,m}=f_{n,m}\Psi_{n,m},\quad \Psi_{n,m+1}=L^{(2)}_{n,m}\Psi_{n,m},\label{lax_d_dress_N}}
where 
$$f_{n,m}=L^{(1)}_{n+N-1,m}L^{(1)}_{n+N-2,m}\dots L^{(1)}_{n,m}.$$

We rewrite the first equation of (\ref{lax_d_dress_N}) in the special form (\ref{lin_equation2}). To this end  the matrix $L^{(1)}_{n,m}$ is factorized as follows:
$$L^{(1)}_{n,m}=\delta_{n,m}Z\rho_{n,m},$$
where $Z=\diag(1,\lambda+\frac{1}{4\alpha_n})$ and
$$\delta_{n,m}=\matrixx{1&0\\ -\frac{2}{u_{n,m}-1}&1}, \quad \rho_{n,m}=\matrixx{1&2\lambda\alpha_n(u_{n,m}+1)\\ 0&\frac{u_{n,m}+1}{u_{n,m}-1}(1+4\lambda\alpha_n)}.$$
Then we pass to $\Psi_{n,m}=\rho^{-1}_{n,m}\Phi_{n,m}$ in the linear system (\ref{lax_d_dress_N}) and finally get \eq{\Phi_{n+N,m}=P_{n,m}Z\Phi_{n,m},\quad \Phi_{n,m+1}=R_{n,m}\Phi_{n,m},  \label{lax_h1_NP2}}
where $P_{n,m}=\rho_{n+N,m}  L^{(1)}_{n+N-1,m}L^{(1)}_{n+N-2,m}\dots L^{(1)}_{n+1,m} \delta_{n,m}$ and $R_{n,m}=\rho_{n,m+1}L^{(2)}_{n,m}\rho^{-1}_{n,m}$.

 It is easy to check that $P_{n,m}(\lambda)$ satisfies all the conditions of Theorem \ref{diagon}, as it is analytical around the point $\lambda=-\frac{1}{4\alpha_n}$. The leading principal minors $\det P_{n,m}({\bf u},\lambda)$ and $\det_1 P_{n,m}({\bf u},\lambda)$ at $\lambda=-\frac{1}{4\alpha_n}$ are rational functions of ${\bf u}=(u_{n},u_{n+1},...,u_{n+N})\in{\C}^{N+1}$ and are not identically zero. Therefore they do not vanish in a domain in ${\C}^{N+1}$.
So, as in preceding examples, there exists a hierarchy of conservation laws. Another hierarchy for eq. (\ref{d_dress}) exists  with no  restriction on the coefficient $\alpha_n$. It has been constructed in \cite{HY13} by diagonalization of the second equation of the $L-A$ pair \eqref{lax_d_dress}.

\subsection{Construction of generalized symmetries}\label{sec_sym}

Results of the method of formal diagonalization can be successfully used for the calculation of generalized symmetries together with their $L-A$ pairs, and we explain here how to do that.
Then in Section \ref{ex_sym} we apply this procedure to the H1 equation  (\ref{abs_h1}, \ref{h1_con}, \ref{per_h1},\ref{noteq}) and to the dressing chain (\ref{dress},\ref{per_dress}). 

We present a scheme applicable to all the three equations under consideration as well as to analogous ones. We describe it by example of the first operator  of the $L-A$ pairs  (\ref{lax_h1},\ref{lax_dress},\ref{lax_d_dress}), and this operator is denoted  here by $L_n^{(1)}$, as the index $m$ is inessential. The corresponding composite operator, we will use, is \eq{\label{Lcomp}L=D_n^{-N}f_n,\quad f_n=L^{(1)}_{n+N-1}L^{(1)}_{n+N-2}\dots L^{(1)}_{n}.} 

Let us denote by $ T_k$ a formal series which diagonalizes the operator $ L$ in the  neighborhood of $\alpha_{n+k}$ with $0\leq k\leq N-1$. Then \eq{L_{0,k}= T_k^{-1} L T_k\label{L0diag}} is a diagonal operator with coefficients which are series in powers of $\lambda-\alpha_{n+k}$. 

Let us describe a class of formal series $B_{n+k,K}=\sum_{j=-K}^{\infty}(\lambda-\alpha_{n+k})^jB_{n+k,K,j}$ satisfying the equation 
\begin{equation}\label{bn}
[L,B_{n+k,K}]=0.
\end{equation}
It can be proved \cite{HY13} that $B_{n+k,K}^{(0)}=T_k^{-1}B_{n+k,K}T_k$ is a formal series with diagonal coefficients: 
\begin{equation}\label{b0_cond}
B_{n+k,K}^{(0)}=\sum_{j=-K}^{\infty}(\lambda-\alpha_{n+k})^jB_{n+k,K,j}^{(0)}, \quad \mbox{such that} \quad D_n^{N}B_{n+k,K,j}^{(0)}=B_{n+k,K,j}^{(0)}.
\end{equation}
The converse is also true, namely, the series $B_{n+k,K}=T_kB_{n+k,K}^{(0)}T_k^{-1}$ solves the equation (\ref{bn}) for any $B_{n+k,K}^{(0)}$ satisfying (\ref{b0_cond}). 

Notice that the operator $L$ can be diagonalized around any of the singular points $\alpha_0,\, \alpha_{1},\,...,\alpha_{N-1}$, and thus a formal series $B_{n+k,K}$ commuting with $L$ can be constructed around any of these points. Therefore there is a class of objects of the form
$B_K=\sum_{k=0}^{N-1}B_{n+k,K}$
solving the equation $[L,B_K]=0$. Cutting off simultaneously all the infinite sums in $B_K$ and finding an appropriate $X_K$ which does not depend on $\lambda$, we construct a rational function:
\eq{A_n^{(K)}=\sum_{k=0}^{N-1}\sum_{j=-K}^{-1}(\lambda-\alpha_{n+k})^jB_{n+k,K,j}+X_K, \quad K\geq 1.\label{AK}}
It is remarkable that this function can be constructed in such a way that the operator equation 
\begin{equation}\label{combined_Lax}
\frac{d}{dt}L=[L,A_n^{(K)}]
\end{equation}
defines a generalized symmetry of equations under consideration. 

We note that there is no general algorithm of finding the operator $X_K$ defining $A^{(K)}_n$ of \eqref{AK}. However in examples below, in Section \ref{ex_sym}, we manage to find it. 

 
The first equation of an $L-A$ pair obtained is rather complicated due to the structure of $L$:
\begin{equation}\label{comp_Lax_pair}
L\Psi_n=\Psi_n, \quad \Psi_{n,t}=A^{(K)}_n\Psi_n.
\end{equation}
Let us explain now why it can be drastically simplified. 

Due to the formula \eqref{Lcomp} for $L$, eq. (\ref{combined_Lax}) is represented as: 
\eqs{\label{combin_Lax2}
\frac{d}{dt}f_n=f_nA^{(K)}_n-A^{(K)}_{n+N}f_n.
}
We denote $W=L^{(1)}_{n+N-2}L^{(1)}_{n+N-3}\dots L^{(1)}_{n}$, then this equation takes the form
$$\frac{dL^{(1)}_{n+N-1}}{dt}W+L^{(1)}_{n+N-1}\frac{dW}{dt}=L^{(1)}_{n+N-1}WA^{(K)}_n-A^{(K)}_{n+N}L^{(1)}_{n+N-1}W.
$$
Evidently it implies
$$(L^{(1)}_{n+N-1})^{-1}(\frac{d}{dt}L^{(1)}_{n+N-1}+A^{(K)}_{n+N}L^{(1)}_{n+N-1}-L^{(1)}_{n+N-1}A^{(K)}_{n+N-1})=$$
$$=-(\frac{d}{dt}W+A^{(K)}_{n+N-1}W-WA^{(K)}_{n})W^{-1}=C_{n+N-1},$$
and we have
\begin{equation}\label{Ln}
\frac{d}{dt}L^{(1)}_{n}=L^{(1)}_{n}A^{(K)}_n-A^{(K)}_{n+1}L^{(1)}_{n}-L^{(1)}_{n}C_n.
\end{equation}

In order to specify the structure of $C_n$, let us deduce for it an equation. To this end we replace the derivatives at the left hand side of (\ref{combin_Lax2}) by using eq. (\ref{Ln}). After some elementary simplifications, one gets
\begin{eqnarray}\nonumber
C_{n+N-1}L^{(1)}_{n+N-2}L^{(1)}_{n+N-3}\dots L^{(1)}_{n}+L^{(1)}_{n+N-2}C_{n+N-2}L^{(1)}_{n+N-3}\dots L_{n}^{(1)}+\dots\\
+L^{(1)}_{n+N-2}L^{(1)}_{n+N-3}\dots L^{(1)}_{n}C_{n}=0. \label{equat_K}
\end{eqnarray}
Calculating the operator $\Omega_{n+1} L_n^{(1)}-L^{(1)}_{n+N-1}\Omega_n$, where $\Omega_n$ is the left hand side of  \eqref{equat_K}, we are led to the relation $f_nC_n-C_{n+N}f_n=0$  which is equivalent to $[L,C_n]=0$. By applying the conjugation \eqref{L0diag}, we find $[L_{0,k},C_{0,n,k}]=0$, where $C_{0,n,k}=T_k^{-1}C_nT_k$ is a diagonal matrix such that $D_n^NC_{0,n,k}=C_{0,n,k}$. 
On the other hand, by construction, $C_n$ is a rational matrix function of $\lambda$ depending on a finite number of the dynamical variables. According to the standard linear algebra theory, it can be diagonalized by applying a conjugation matrix $R$ which also depends, unlike $T_k$, on a finite set of the dynamical variables: $R^{-1}C_nR=C_{0,n,k}$. 

 
Let us prove that $C_{0,n,k}$ is a scalar matrix, i.e. it is proportional to the unity matrix. Suppose the contrary, then the commutativity relation $C_{0,n,k}R^{-1}T_k=R^{-1}T_kC_{0,n,k}$ implies that the product $\hat T_k=R^{-1}T_k$ is a diagonal matrix, therefore $T_k=R\hat T_k$. The relation $L_{0,k}=T_k^{-1}LT_k$ implies the equation  $R^{-1}LR=\hat T_kL_{0,k}\hat T_k^{-1}$, which shows that the operator $L$ is diagonalized by a matrix $R$ depending on a finite set of dynamical variables. This contradicts Theorem \ref{diagon}, and thus $C_{0,n,k}=C_n=c(n,\lambda)E$ is a scalar matrix. The equation (\ref{equat_K}) immediately gives 
\begin{equation}\label{condit_K}
c(n,\lambda)+c(n+1,\lambda)+\dots +c(n+N-1,\lambda)=0.
\end{equation}
It follows from the equation $[L,c(n,\lambda)]=0$ equivalent to $c(n,\lambda)=c(n+N,\lambda)$ that $c(n,\lambda)$ does not depend on the dynamical variables $u_{n+j}$.  

Evidently eq. (\ref{Ln}) can be reduced to the usual Lax equation. Indeed, fixing the value $n=n_0$ and setting $\hat A^{(K)}_n=A^{(K)}_n+\sum_{k=n_0}^{n-1}c(k,\lambda)$, we get the equation
\begin{equation}\label{final_lax}
\frac{d}{dt}L^{(1)}_{n}=L^{(1)}_{n}\hat A^{(K)}_n- \hat A^{(K)}_{n+1}L^{(1)}_{n}.
\end{equation}
This equation is nothing but the compatibility condition for the linear system
$$\Psi_{n+1}=L^{(1)}_{n}\Psi_{n},\quad \frac{d}{dt}\Psi_{n}=\hat A^{(K)}_n\Psi_{n}.$$
Let us emphasize that due to eq. (\ref{condit_K}) the term $\sum_{k=n_0}^{n-1}c(k,\lambda)$ contains not more than $N-1$ summands.

We hypothesize that the function $c(n,\lambda)$ is always zero, as in all examples below, but we cannot prove this fact. 


\section[Examples of generalized symmetries and their L-A pairs]{Examples of generalized symmetries and their $L-A$ pairs}\label{ex_sym}
In this section we apply the diagonalization procedure of Section \ref{sec_sym} to the H1 equation (\ref{abs_h1},\ref{h1_con}) and the dressing chain (\ref{dress}). In some cases the generalized symmetries are given in Sections \ref{sec_h1},\ref{sec_dr}, and we just construct corresponding $L-A$ pairs. In one case we find a new generalized symmetry together with its $L-A$ pair.

\subsection{H1 equation}
Let us consider the H1 equation (\ref{abs_h1},\ref{h1_con}) with an arbitrary coefficient $\beta_m$ and a periodic $\alpha_n$ satisfying the restriction \eqref{per_dress}.
Following Section \ref{sec_sym}, we construct $L-A$ pairs of the form:
\eq{\Psi_{n+1,m}=L^{(1)}_{n,m}\Psi_{n,m},\quad \frac{d}{dt_N}\Psi_{n,m}=A_{n,m}^{(1,N)}\Psi_{n,m}\label{lax_sym}.}
Here $A_{n,m}^{(K,N)}$ corresponds to $\hat A_{n}^{(K)}$ of the previous section, $N$ is the period of $\alpha_n$, and $K=1$ indicates the lowest term of a hierarchy of generalized symmetries. It turns out that the additional term $X_K$ of \eqref{AK} is equal to zero here.

In case of the period $N=1$, i.e. when $\alpha_n$ is a constant, the matrix $A^{(1,1)}_{n,m}$ reads:
\seq{A_{n,m}^{(1,1)}=\frac{1}{(\lambda-\alpha_n)(u_{n+1,m}-u_{n-1,m})}\matrixx{u_{n-1,m}&1\\-u_{n+1,m}u_{n-1,m}&-u_{n+1,m}}.} The compatibility condition for the $L-A$ pair \eqref{lax_sym} is equivalent in this case to the generalized symmetry \eqref{sym_nond}, and this is one more way to construct this symmetry.

In case of the period $N=2$, we obtain for the generalized symmetry \eqref{sym_h1_2} an $L-A$ pair of the form \eqref{lax_sym} defined by the following matrix:
\seqs{A_{n,m}^{(1,2)}=&\frac{c_n}{(\lambda-\alpha_n)(\gamma_n+(u_{n,m}-u_{n-2,m})(u_{n-1,m}-u_{n+1,m}))}\\\cdot&\matrixc{u_{n-2,m}-u_{n,m}\\\gamma_n+u_{n-1,m}(u_{n,m}-u_{n-2,m})}\times\matrixx{u_{n+1,m}\,,&1}\\+&\frac{c_{n+1}}{(\lambda-\alpha_{n+1})(\gamma_n+(u_{n,m}-u_{n+2,m})(u_{n-1,m}-u_{n+1,m}))}\\&\cdot\matrixc{1\\-u_{n-1,m}}\times\matrixx{\gamma_n+u_{n+1,m}(u_{n,m}-u_{n+2,m})\,,&u_{n+2,m}-u_{n,m}},} where by $\times$ we denote the matrix product.

\subsection{Dressing chain }

Here we consider the dressing chain \eqref{dress} with a periodic $\alpha_n$ satisfying the restriction \eqref{per_dress}.
As in previous examples, we construct $L-A$ pairs of the form
\eq{\Psi_{n+1}=L^{(1)}_{n}\Psi_{n},\quad \frac{d}{dt_N}\Psi_{n}=A_{n}^{(1,N)}\Psi_{n}\label{lax_sym_dr},}
where $N$ is the period of $\alpha_n$.
It turns out that, in case of the dressing chain, the additional term $X_K$ of \eqref{AK} is a matrix with a nonzero element in the lower left corner only.

In the case $N=2$ we have found in Section \ref{sec_or_dr} the generalized symmetry \eqref{sym_dr_2}.
Putting $c=0$ to exclude from consideration the master symmetry, we construct the $L-A$ pair \eqref{lax_sym_dr} defined by the following matrix:
\seqs{A_{n}^{(1,2)}&=\frac{b_{n}}{(\lambda+\alpha_n)(\gamma_n+v_{n-1}v_{n-2})}\left(\matrixx{v_{n-2}\\-\gamma_n-v_{n-2}u_{n-1}}\times\matrixx{u_n,&-1}\right.\\&+\left.(\lambda+\alpha_n)\matrixx{0&0\\-v_{n-2}&0}\right)\\
&+\frac{b_{n+1}}{(\lambda+\alpha_{n+1})(\gamma_n-v_{n}v_{n-1})} \left(\matrixx{1\\-u_{n-1}}\times\matrixx{\gamma_n-u_nv_n,&v_n}\right.\\&+\left.(\lambda+\alpha_{n+1})\matrixx{0&0\\v_{n}&0}\right),}
where $v_n=u_{n+1}+u_n$.

In the case  $N=3$, let us denote by $a_n$ an arbitrary three-periodic function, so that $a_{n+3}\equiv a_n$, and let us introduce the notations
\seqs{\gamma_n=\alpha_{n+1}-\alpha_n,\quad v_{n}=u_{n+1}+u_n,\quad U_{n}=v_{n+1}v_{n}+\gamma_n,\\V_{n}=v_{n+1}v_{n}v_{n-1}+\gamma_{n+2}v_{n+1}-\gamma_{n+1}v_{n-1},}
where $\gamma_n\neq0$ for any $n$ due to \eqref{per_dress}.
We construct a matrix $A_{n}^{(1,3)}$ determining the $L-A$ pair \eqref{lax_sym_dr}, which can be expressed as follows:
\seqs{A_{n}^{(1,3)}=&\frac{a_n}{(\lambda+\alpha_n)V_{n-2}}A_n+\frac{a_{n+1}}{(\lambda+\alpha_{n+1})V_{n-1}}B_n+\frac{a_{n+2}}{(\lambda+\alpha_{n+2})V_{n}}C_n,\\
A_n=&\left(\begin{array}{c}U_{n-3}\\U_{n-3}u_n-V_{n-2}\end{array}\right)\times (u_n,\quad -1)+(\lambda+\alpha_n)\left(\begin{array}{cc}0&0\\-U_{n-3}&0\end{array}\right),\\
B_n=&\left(\begin{array}{c}v_{n-2}\\v_{n-2}u_n-U_{n-2}\end{array}\right)\times (v_nu_n-\gamma_n,\quad -v_n)+(\lambda+\alpha_{n+1})\left(\begin{array}{cc}0&0\\-v_{n-2}v_n&0\end{array}\right),\\
C_n=&\left(\begin{array}{c}1\\-u_{n-1}\end{array}\right)\times \big((U_n+\gamma_{n+2})u_n+\gamma_{n+2}v_{n+1},\quad -U_{n}-\gamma_{n+2}\big)\\&+(\lambda+\alpha_{n+2})\left(\begin{array}{cc}0&0\\-U_{n}-\gamma_{n+2}&0\end{array}\right).}

The compatibility condition for this $L-A$ pair is equivalent to the equation
\seq{\frac{du_n}{dt_3}=\frac{U_{n+1}+\gamma_n}{V_{n+1}}a_n-\frac{U_{n-3}}{V_{n-2}}a_n+\frac{U_{n-2}-v_{n}v_{n-2}}{V_{n-1}}a_{n+1}+\frac{v_{n+1}v_{n-1}-U_n-\gamma_{n+2}}{V_n}a_{n+2}.} 
This is a new third order generalized symmetry of the dressing chain with the three-periodic coefficient $\alpha_n$. This symmetry is the lowest term of a hierarchy in this case.

\section{Examples of conservation laws}\label{sec_claws}
In this section we apply the diagonalization procedure of Section \ref{sec_d_law} to three equations under consideration with periodic coefficients and write down for them a number of conservation laws. 
The structure of those conservation laws essentially differs from the standard one, cf. \cite{HY13,gmy14,gy14}.

\subsection{Dressing chain}

We consider the dressing chain \eqref{dress} with a two-periodic coefficient $\alpha_n$ satisfying \eqref{p2n11}. 
In case of an arbitrary $\alpha_n$, one hierarchy of conservation laws has been constructed in \cite{HY13}. Conserved densities in that hierarchy depend on  the $x$-derivatives of $u_n$, and those conservation laws can be called ones in the $x$-direction. Here we construct some conservation laws in the $n$-direction, which can be represented in the form:
\seq{D_x p^{(j)}_n=(1-D_n^2)q_n^{(j)},\quad j\geq0.}
The functions $p^{(j)}_n$ and $q_n^{(j)}$ depend on the shifts of $u_n$, and first two pairs of them read:
\eqs{&p_n^{(0)}=\log(v_{n+1}v_n+\gamma_n),\qquad\qquad\qquad q^{(0)}_n=u_n,\\
&p_n^{(1)}=\frac{(v_{n+1}+v_{n-1})v_{n-2}+\gamma_n}{(v_{n+1}v_{n}+\gamma_n)(v_{n-1}v_{n-2}+\gamma_n)},\quad
q_n^{(1)}=\frac{v_{n-2}}{v_{n-1}v_{n-2}+\gamma_n},\label{laws_dr}} where
\seq{v_{n}=u_{n+1}+u_{n},\quad \gamma_n=\alpha_{n+1}-\alpha_n,} and $\gamma_n\neq0$ for all $n$ due to \eqref{p2n11}.

The conserved density $p^{(0)}_n$ depends on three variables, while $p^{(1)}_n$ depends on five ones. The following conditions take place:
\seq{\frac{\partial^2 p^{(0)}_n}{\partial u_n\partial u_{n+2}}\neq0,\qquad \frac{\partial^2 p^{(1)}_n}{\partial u_{n-2}\partial u_{n+2}}\neq0} for all $n$. In accordance with a general theory of \cite{ly97}, the number of variables of such functions cannot be reduced. This shows, in particular, that two conservation laws defined by \eqref{laws_dr} are essentially different. These conservation laws can be called the three- and five-point ones, respectively.

\subsection{Discrete dressing chain}
Let us consider the discrete dressing chain \eqref{d_dress} with a two-periodic \eqref{p2n13} coefficient $\alpha_n\neq0$ for any $n$. 
In case of an arbitrary $\alpha_n$, one hierarchy of conservation laws has been constructed in \cite{HY13}. Conserved densities in that hierarchy depend on the $m$-shifts of $u_{n,m}$, and those conservation laws can be called ones in the $m$-direction. Here we construct some conservation laws in the $n$-direction, and those can be represented in the form:
\eq{(D_m-1)p^{(j)}_{n,m}=(D_n^2-1)q^{(j)}_{n,m},\quad j\geq0.}
The functions $p^{(j)}_{n,m}$ and $q_{n,m}^{(j)}$ will depend on the $n$-shifts of $u_{n,m}$.

We use the notation
\seqs{U_{n,m}=(u_{n,m}+u_{n-1,m})(u_{n-1,m}+u_{n-2,m})+(\beta_n-1)(u_{n-1,m}^2-1),} where $\beta_n=\alpha_{n+1}/\alpha_n\neq1$ for any $n$ due to \eqref{p2n13}.
Two simplest conservation laws are given by:
\eq{p^{(0)}_{n,m}=\log{\frac{U_{n+2,m}}{(u_{n+1,m}-1)(u_{n,m}-1)}},\quad q^{(0)}_{n,m}=\log(u_{n,m+1}-1);}
\eqs{p^{(1)}_{n,m}=&\frac{\beta_n(u_{n+1,m}^2-1)}{U_{n+2,m}}+\frac{(u_{n,m}^2-1)(u_{n+2,m}+u_{n+1,m})(u_{n-1,m}+u_{n-2,m})}{U_{n+2,m}U_{n,m}},\\ q^{(1)}_{n,m}=&\frac{(u_{n,m}+1)(u_{n-1,m}+u_{n-2,m})}{U_{n,m}}.}

The following conditions are satisfied for all $n,m$:
\seq{\frac{\partial^2 p^{(0)}_{n,m}}{\partial u_{n,m}\partial u_{n+2,m}}\neq0,\qquad \frac{\partial^2 p^{(1)}_{n,m}}{\partial u_{n-2,m}\partial u_{n+2,m}}\neq0.} In accordance with some theoretical remarks of \cite{gmy14}, the number of variables of such functions cannot be reduced, in particular, two above conservation laws are essentially different three- and five-point ones.

\subsection{Asymmetric H1 equation}

We construct here conservation laws for the asymmetric example of Section \ref{asH1}, i.e. for the H1 equation with three-periodic coefficient $\alpha_n$ and two-periodic coefficient $\beta_m$. Recall that, in the autonomous case, conservation laws for the H1 equation have been found in \cite{rh07_1}. 

The diagonalization procedure in the neighborhood of the singular point $\alpha_n$ gives us a hierarchy of conservation laws of the form
\eq{(D_m-1)p^{(j)}_{n,m}=(D_n^3-1)q^{(j)}_{n,m},\quad j\geq0.}
First two of them are defined by the following functions:
\eqs{p^{(0)}_{n,m}=\log V_{n-1,m},\quad q^{(0)}_{n,m}=\log(u_{n,m+1}-u_{n-1,m});}
\eqs{p^{(1)}_{n,m}=\frac{1}{V_{n,m}}\left(v_{n+1,m}+\frac{U_{n+2,m}}{v_{n,m}}\left(\frac{U_{n,m}}{\gamma_n}-\frac{\gamma_{n+2}v_{n-2,m}}{V_{n-3,m}}\right)\right),\\ q^{(1)}_{n,m}=\frac{u_{n+1,m}-u_{n,m+1}}{\alpha_n-\beta_m}\left(\frac1{v_{n,m}}-\frac{\gamma_{n+2}}{U_{n-1,m}}\left(\frac1{v_{n,m}}+\frac{\gamma_n}{V_{n-3,m}}\right)\right).}
Here we use the notations
\seqs{U_{n,m}=v_{n+1,m}v_{n,m}+\gamma_n,\\ V_{n,m}=v_{n+3,m}v_{n+2,m}v_{n+1,m}-\gamma_nv_{n+3,m}+\gamma_{n+2}v_{n+1,m},}
where $v_{n,m}$ and $\gamma_n$ are given in eq. \eqref{sn32}. 

Standard relations take place for all $n,m$: 
\seq{\frac{\partial^2 p^{(0)}_{n,m}}{\partial u_{n-1,m}\partial u_{n+3,m}}\neq0,\qquad \frac{\partial^2 p^{(1)}_{n,m}}{\partial u_{n-3,m}\partial u_{n+4,m}}\neq0,} i.e. these conservation laws are essentially different.
We have the five- and eight-point conservation laws in the $n$-direction, the simplest of those we can get by this procedure. 

The diagonalization procedure in the neighborhood of the singular point $\beta_m$ allows one to construct the second hierarchy of conservation laws of the form
\eq{(D_n-1)\hat q^{(j)}_{n,m}=(D_m^2-1)\hat p^{(j)}_{n,m},\quad j\geq0.}
First two conservation laws are defined by:
\eq{\hat q^{(0)}_{n,m}=\log(w_{n,m+1}w_{n,m}+\delta_m),\quad \hat p^{(0)}_{n,m}=\log(u_{n+1,m}-u_{n,m-1});}
\eqs{\hat q^{(1)}_{n,m}=\frac{w_{n,m-1}(w_{n,m+2}+w_{n,m})-\delta_m}{(w_{n,m+2}w_{n,m+1}-\delta_m)(w_{n,m}w_{n,m-1}-\delta_m)},\\\hat p^{(1)}_{n,m}=\frac{w_{n,m-1}(u_{n+1,m}-u_{n,m+1})}{(\alpha_n-\beta_m)(w_{n,m}w_{n,m-1}-\delta_m)},}
where $w_{n,m}$ and $\delta_m$ are given in \eqref{sm32}.

Standard relations are satisfied for all $n,m$: \seq{\frac{\partial^2 \hat q^{(0)}_{n,m}}{\partial u_{n,m-1}\partial u_{n,m+2}}\neq0,\qquad \frac{\partial^2 \hat q^{(1)}_{n,m}}{\partial u_{n,m-2}\partial u_{n,m+3}}\neq0,} i.e. we have four- and six-point conservation laws in the $m$-direction. 

We can see that the structure of conservation laws in different directions is quite different as well as one of generalized symmetries.

\section{The nature of generalized symmetries}\label{sec_sys}
In this section we briefly discuss the nature of second order generalized symmetries obtained in Sections \ref{sec_h1}-\ref{sec_ddr}. Recently, some examples of discrete equations of the form \eqref{gF} have been obtained, whose simplest generalized symmetries in at least one of directions are of the second order as well. Most of those generalized symmetries \cite{a11,mx13,shl14} are similar  to the Ito-Narita-Bogoyavlensky lattice equation. In one of such examples, the second order generalized symmetry is of the relativistic Toda type \cite{gy12,gmy14}. The second order symmetries obtained in this paper are of the relativistic Toda type too.

The discrete-differential nonautonomous scalar equations with discrete periodic coefficients can be rewritten as autonomous systems, see \cite{ly97}.
For example, eq. \eqref{sym_h1_2} for any fixed value of $m$ can be represented as an autonomous system of two equations, as it has the two-periodic coefficients $c_n$ and $\gamma_n$. Introducing the notations
\eq{v_k=u_{2k,m},\quad w_k=u_{2k+1,m},\quad \gamma=\gamma_{2k}\neq0,\quad A=c_{2k},\quad B=c_{2k+1},} we obtain the following system:
\eqs{\dot v_k=\frac{A(v_k-v_{k+1})}{\gamma+(w_{k-1}-w_k)(v_k-v_{k+1})}+\frac{B(v_k-v_{k-1})}{\gamma+(w_{k-1}-w_k)(v_k-v_{k-1})},\\ \dot w_k=\frac{A(w_{k-1}-w_{k})}{\gamma+(w_{k-1}-w_k)(v_{k}-v_{k+1})}+\frac{B(w_{k+1}-w_{k})}{\gamma+(v_{k}-v_{k+1})(w_{k+1}-w_{k})}.\label{svw}}

Here we have two similar and commuting with each other systems: the first one is defined by $A=1,\ B=0$ and the second one by $A=0,\ B=1$. According to their symmetry structure, such systems are similar to relativistic Toda type systems, cf.  \cite[Section 5.1]{asy00}. The system \eqref{svw} is an analogue of the well-known Ablowitz–Ladik example which is a linear combination of two commuting systems of equations of the relativistic Toda type (see, e.g., Section 5.2 in \cite{asy00}). Two other generalized symmetries \eqref{sym_dr_2} and \eqref{sym_ddr_2} with $c=0$ are of the same kind. The case  $c\neq0$ is not periodic and corresponds to the master symmetry.

In case of the system \eqref{sym_h1_2}, we can illustrate the same property in a more explicit way. Let us consider the system \eqref{svw} with $A=1$ and $B=0$. It can be checked by direct calculation that each of the functions $v_k$ and $w_k$ satisfies, up to rescaling the time, the following lattice equation: \eq{\ddot U_k=\dot U_k^2\left(\frac{\dot U_{k-1}}{(U_k-U_{k-1})^2}-\frac{\dot U_{k+1}}{(U_k-U_{k+1})^2}+\frac1{U_k-U_{k-1}}+\frac1{U_k-U_{k+1}}\right)\label{rel_T}.}  This is the well-known equation of the relativistic Toda type, see e.g. the review articles \cite[Section 4.2]{asy00}  and \cite[Section 3.3.3]{y06}. The same is true for the system \eqref{svw} with $A=0$ and $B=1$.

Note that any solution $u_{n,m}$ of the second order symmetry \eqref{sym_h1_2} is transformed into a solution $\hat u_{n}$ of an equation of the form \eqref{sym_dr_2} by the following formula: \seq{\hat u_{n}= u_{n+1,m}-u_{n,m}.} More precisely, the function $\hat u_{n}$ satisfies eq. \eqref{sym_dr_2} with $c=0$ and slightly changed $\gamma_n$ and $b_n$. This shows that the symmetries   \eqref{sym_dr_2} with $c=0$ and \eqref{sym_h1_2} are almost the same and have the same nature.

\section{Conclusions}

In Sections \ref{sec_h1}-\ref{sec_ddr} we have proved a number of theorems which allow us to formulate the following hypothesis: 

{\bf Hypothesis 1.} {\it The generalized symmetries of eqs. (\ref{abs_h1},\ref{dress},\ref{d_dress}) in the $n$-direction exist if and only if the coefficient $\alpha_n$ is periodic. If $\alpha_n$ has the period $N$, then the simplest generalized symmetries of these equations have the order $N$.}

As for conservation laws, we assume that a similar picture takes place: 

{\bf Hypothesis 2.} {\it A hierarchy of conservation laws for eqs. (\ref{abs_h1},\ref{dress},\ref{d_dress}) in the $n$-direction exists if and only if the coefficient $\alpha_n$ is periodic.}

The case of the $m$-direction for eq. \eqref{abs_h1} is analogous.
The first hypothesis is substantiated in Sections \ref{sec_h1}-\ref{sec_ddr} in case of the first and second order  generalized symmetries. For equations under consideration, which have periodic coefficients with an arbitrarily large period, both hypotheses are partially confirmed in Section \ref{sec_theory}. 

In this section we develop a theory for the case of nonautonomous discrete equations, which allows one, in particular, to construct generalized symmetries and conservation laws for eqs. (\ref{abs_h1},\ref{dress},\ref{d_dress}) with periodic coefficients. In Sections \ref{ex_sym},\ref{sec_claws} we apply this theory to construct some examples. 

The picture for all nonautonomous equations of the Adler-Bobenko-Suris list should be the same as for eq. \eqref{abs_h1}.

We also come to an opinion that eqs. (\ref{abs_h1},\ref{dress},\ref{d_dress}) with periodic coefficients are integrable in the same sense as the autonomous equations possessing $L-A$ pairs. The case of nonperiodic coefficients seems to be much more difficult from the standpoint of integrability.

\paragraph{Acknowledgments.}
This work has been supported by the Russian Foundation for Basic Research (grant numbers: 13-01-00070, 14-01-97008-r-povolzhie-a).

\end{document}